\documentclass[fleqn,twoside]{article}
\usepackage{espcrc2, verbatim}
\usepackage{amssymb,latexsym,amsmath,epsfig}

\newcommand{\sun}{\odot}
\newcommand{\apj}{ApJ}

\newcommand{\mnras}{MNRAS}
\newcommand{\aap}{A\&A}
\newcommand{\araa}{ARA\&A}
\newcommand{\nat}{Nature}
\newcommand{\iaucirc}{IAU circ.}

\title{Thin Disk Models of Anomalous X-ray Pulsars}

\author{K. Yavuz Ek\c{s}i\address[MCSD]{Sabanc\i\ University, Orhanl\i--Tuzla, \.{I}stanbul 34956, Turkey}
        and
        M.Ali Alpar\addressmark[MCSD] }

\begin{document}

\begin{abstract}
We discuss the options of the fall-back disk model of
Anomalous X-Ray Pulsars (AXPs). We argue that the power-law index of the mass 
inflow rate during the propeller stage can be lower than those employed 
in earlier models. We take into account the effect of the super-critical mass 
inflow at the earliest stages on the inner radius of the disk and argue that
the system starts as a propeller. 
Our results show that, assuming a fraction of the mass inflow is
accreted onto the neutron star, the fall-back disk scenario can produce AXPs
for acceptable parameters.
\vspace{1cm}
\end{abstract}

\maketitle

\section{Introduction}

The models for explaining the X-ray energy output of Anomalous X-ray Pulsars
(AXPs, see \cite{mereghetti} for a review) can be classified in two groups.
Magnetar models \cite{duncan92,thompson95,thompson96} assign the persistent emission 
and bursts \cite{AXPburst1,AXPburst2} to the extra-ordinary magnetic 
field of the neutron star while accretion models 
\cite{MS95,vanParadijs,CHN,alpar01,marsden01,EA03} 
employ neutron stars with intrinsic properties similar to other neutron stars and 
try to explain the X-ray emission and spin properties by the way matter is supplied.
AXPs are very similar to Soft Gamma Ray Repeaters (SGRs) \cite{hurley,kouve98,kouve99}
and probably have the same nature.

Magnetar model is successful in
modelling the SGR/AXP bursts and spin-down rates but can not explain
the period clustering \cite{PM02} except in one set of field decay models
under special conditions \cite{CGP}. Interaction of the neutron star with
a fall-back disk towards equilibrium can naturally account for the period 
clustering.

Mereghetti \& Stella (1995)\cite{MS95} proposed that AXPs were accreting from a 
very low mass companion. There are severe upper limits for
the mass of such a companion and no orbital motion signatures have been detected.
Van Paradijs et al. (1995) \cite{vanParadijs} proposed a fossil disk model in which 
the disk results from the final spiral-in destruction of the binary. 
Such a model addresses the absence of the binary companion but can not
account for the young ages of AXPs indicated by the supernova remnant (SNR)
associations \cite{gaensler,tagieva}. Fall-back disk models 
\cite{CHN,alpar01,marsden01,EA03} can address the young ages of AXPs
and the absence of a mass donor companion. In all fall-back disk models
the spin down of the neutron star is achieved by the propeller effect
\cite{shv70,PR72,IS75,fabian75,DP81,ihsan01,ihsan02} in which the rapid
rotation of the neutron star prevents inflowing matter from
accreting onto the surface of the star. Even when the star is slow enough
that accretion can commence, it can continue to spin-down 
until torque equilibrium. AXPs must be those spinning-down accreting systems
near torque equilibrium.

Recent detections in the infrared (and optical) support the
existence of a disk \cite{israel03}. The magnetar models do not
have a prediction for infrared emission. 
Fall-back disks are not necessarily geometrically thin but,
as the only models available  for calculation, they
warrant a careful study to decide whether interaction with a
fall-back disk can produce neutron stars with AXP properties at the
ages indicated by SNR-AXP associations. 
In this work we discuss the options of the fall-back disk model in view 
of the recent disk solution employed by Ek\c{s}i \& Alpar (2003)\cite{EA03}.

\section{Fall-back Disk Models}

Alpar (2001) \cite{alpar01}
proposed a classification of young neutron stars in terms of the
absence or presence and properties of a fall-back disk. According
to this model AXPs, SGRs and Dim Thermal Neutron Stars (DTNs) have
similar periods because they are in an asymptotic spin-down phase
in interaction with a fall-back disk. The different classes
represent alternative pathways of neutron stars. Radio pulsars
have no disks or encounter very low mass inflow rates while Radio
Quiet Neutron Stars (RQNS) have such high mass inflow rates that
their pulse periods are obscured by the dense medium around.
This model employs constant mass inflow rate corresponding to a 
constant equilibrium period. As a fall-back disk is not replenished,
the mass inflow rate of the disk should decline while the disk 
diminishes. The mass flow rates characterizing different evolutionary paths
given in this model are then 
representative values and are to be replaced by the initial mass
of the disk in a time-dependent model.

Chatterjee et al. (2000) (CHN \cite{CHN}) employed a time-dependent 
thin disk model \cite{pringle74,CLG} in which the angular momentum of the disk 
is constant and the mass of the disk declines with a power-law.
Accordingly, the mass inflow rate also declines as a power-law
\begin{equation}
\dot{M}=\dot{M}_0\left(\frac{t}{T_0}\right)^{-\alpha}
\label{Mdot_m}
\end{equation}
where $\dot{M}_0$ is the initial inflow rate and $T_0$ is taken to be 
the dynamical
time scale, $T_d \sim \Omega_K^{-1}$ at the initial inner radius of the disk. 
The power-law index
$\alpha$ depends on the opacity regime. For opacities of
the form $\kappa =\kappa _{0}\rho ^{a}T^{b}$, thin disk equations \cite{FKR92}
can be solved to obtain
\begin{equation}
\alpha =\frac{18a-4b+38}{17a-2b+32}.
\label{alpha}
\end{equation} 
 This yields, $\alpha=19/16$ for electron scattering opacity ($a=b=0$), and $\alpha=5/4$ for 
 bound-free opacity ($a=1$ and $b=-7/2$). 
 CHN assumed electron scattering to be dominating, but employed $\alpha=7/6$ for analytical 
 convenience. 
 As the mass inflow rate declines, the equilibrium period increases and the system
 can never reach a true equilibrium. AXPs are those systems accreting from a 
 diminishing disk while spinning down towards an ever receding
 equilibrium (quasi-equilibrium stage).

Francischelli \& Wijers (2002)\cite{FW} criticized the CHN model, noting that 
whether the disk model can produce
AXPs or not is sensitively dependent on the value of the power-law index 
$\alpha$. They showed that after $\sim 20$ years, the disk is bound-free opacity
dominated and for the corresponding $\alpha$, the model did not produce AXPs with the nominal
disk mass employed by CHN ($M_0=0.006M_{\odot}$). They also
criticized the torque model employed in the CHN model, arguing
that it is too efficient and showed that less efficient torque models
can not produce AXPs (see also \cite{li02}).

While Fracischelli \& Wijers (2002) \cite{FW} is right in arguing that 
$\alpha$ is critical in determining the period
evolution for a single initial disk mass, it is not possible to 
reject the fall-back disk model on this ground.
For every $\alpha$, one 
can find an appropriate initial mass flow rate $\dot{M}_0$ (corresponding to an initial disk mass $M_0$)
that leads to a period evolution in which the system enters the quasi-equilibrium stage 
at $P \cong 5$ s, producing an AXP. The time at which the transition to the quasi-equilibrium stage occurs, 
$T_{tr}$, and the mass flow rate at 
this moment, $\dot{M}_{tr}$, remains nearly the same for each $\alpha$ and its corresponding $\dot{M}_0$. 
One can obtain AXPs with the CHN model 
by employing $\alpha=5/4$ corresponding to bound-free opacity as long as one assumes 
an initial disk mass of $M_0\cong 0.069M_{\odot}$. This larger initial disk mass does not lead to 
higher luminosity at the accretion stage because the disk decays more rapidly with the corresponding $\alpha$.

Ek\c{s}i \& Alpar (2003) (EA03 \cite{EA03}) employed another time-dependent thin disk 
solution \cite{pringle74,pringle91} for the evolution of the disk during the propeller stage.
In this solution mass of the disk 
is constant while the angular momentum of the disk increases
due to the torque 
\begin{equation}
\dot{J}=\dot{J}_0\left(\frac{t}{T_0}\right)^{-\beta}
\label{Jdot_m}
\end{equation}
at the inner boundary of the disk where
\begin{equation}
\beta =\frac{14a+22}{15a-2b+28}.
\label{beta}
\end{equation}
For electron scattering opacity $\beta =11/14$ and for 
bound-free opacity dominated disks $\beta =18/25$. In this solution
mass flows outwards, but Pringle (1991)\cite{pringle91} numerically
shows solutions in which the mass flows inwards, and a torque as 
given in Eq.~(\ref{Jdot_m}) is
necessary to stop accretion at the inner boundary. Pringle (1981)
\cite{pringle81} mentioned that  
``such a
solution might represent a disc around a magnetized star which is rotating
sufficiently rapidly that its angular velocity exceeds the Keplerian
velocity at the magnetosphere'' - i.e. a propeller.
Assuming
the inner radius $R_m$ scales with the Alfv\'en radius  $R_A\propto \dot{M}^{-2/7}$
and 
referring $\dot{J}\propto \sqrt{GMR_m}\dot{M}\propto \dot{M}^{6/7}$, 
one can calculate
an effective $\alpha$ for the propeller regime. For electron-scattering opacity
one obtains $\alpha_p=11/12$ and for bound-free opacity $\alpha_p=21/25$. EA03
used another scaling for determining the inner radius in the propeller regime
$R_m\propto \dot{M}^{-1/5}$ \cite{romanova}.
This implementation for the
inner radius yields $\dot{J}\propto \dot{M}^{9/10}$ and one obtains
$\alpha_p=55/63$ for the electron scattering opacity 
and $\alpha_p=4/5$ for 
the bound-free opacity.

For the sake of consistency with the constant disk mass solution, EA03
assumed that the inflowing matter
flung away by the magnetosphere of the neutron star returns back
to the disk somewhere away from the inner boundary of the disk.
The mass flow rate in the propeller type of solution declines with a softer
power-law index than it does in the accretion type of solution because in the former case
matter cannot accrete and the decrease in the flow rate is only by the
viscous spreading of the disk. Conservation of the mass of the disk during the propeller stage
makes it possible to obtain AXPs with a 50 times smaller initial disk masses
than required by the CHN model. 
Again, this does not mean that mass flow rate during the accretion stage will be smaller 
by this order of magnitude. 

The luminosities of AXPs being less than $L_X \simeq 7\times 10^{35}$ erg s$^{-1}$ imply that 
less than $\dot{M}_X \simeq 4\times 10^{15}$ g s$^{-1}$ of matter is accreting onto these objects.
As Menou et al.\ (2001) \cite{MPH01} have shown, fall-back disks with 
$\dot{M}\lesssim 10^{16}$ g s$^{-1}$ are
neutralized and can not be accreting. This means that AXPs must have mass flow rates in excess 
of $\sim 10^{16}$ g s$^{-1}$ and only a fraction $\eta$ of the inflowing matter should accrete 
onto the neutron star.
In CHN model $\dot{M}_{tr} \gtrsim 10^{16}$ g s$^{-1}$ and
one needs to assume that a fraction $\eta \cong 0.1$ of the inflowing mass 
accretes onto the neutron star, 
in order to explain the observed luminosities of 
$L_X \lesssim 10^{36}$ erg s$^{-1}$. 
In EA03 model one needs to 
assume $\eta \sim 0.01$ in order to explain the luminosities
because the system reaches the quasi-equilibrium stage
rather rapidly, at an epoch during which the
mass flow rate is still high. The difference is due to the 
different formulations employed for the magnetic radius of the disk
during the propeller stage which in turn effects the magnitude of the torque
and the time the system reaches the quasi-equilibrium stage.

\section{The Model Equations}

For finding the inner radius, we follow the EA03 model \cite{EA03}.
The system initially is in the super-critical 
propeller regime and the inner radius of the disk is determined by
\begin{eqnarray}
R_{m} &=&( \mu ^{2}\kappa _{es}/8\pi \sqrt{2}c \Omega _{\ast}) ^{1/6}  
\nonumber \\
&=&31R_{\ast }\mu _{30}^{1/3}( P_{0}/15\mbox{ms})^{1/6}  
\label{R_mEA}
\end{eqnarray}
while the corotation radius is
$R_{c}\simeq 10R_{\ast}(P/15\text{ms})^{2/3}$.
The dynamical time scale $T_d\equiv 1/\Omega_K(R_m)$ correspondingly is
$T_d=1.23\times 10^{-2}\mu _{30}^{1/6}(P_{0}/15\text{ms})^{1/4}$
which is one order of magnitude greater than that employed by CHN. Greater $T_0$ 
makes it possible to obtain the same mass flow history (and same $\dot{M}_{tr}$) 
starting with a smaller 
$\dot{M}_0$ by (and smaller $M_0$). The calculations of Pringle (1991)\cite{pringle91}
show that the power-law evolution starts after a viscous time scale rather than
the dynamical time-scale employed in early models.
Considering the 
ambiguity in $T_0$, 
one should not attempt to be very definite on the initial mass of the 
disk in models employing the power-law evolution.

We model the disk torque as
\begin{equation}
\dot{J}_{\ast } =
\sqrt{GMR_{m}} \dot{M} ( 1-\omega _{\ast }/\omega _{eq })
\label{torque}
\end{equation}
where $\omega_{eq}$ is the equilibrium fastness
parameter which we take to be $0.85$ in this work.
As noted by \cite{FW} and \cite{li02}, this is a very efficient
propeller torque model compared to some other available propeller
torques obtained through simple energy or angular momentum
arguments \cite{PR72,IS75,fabian75}. 
Recently Ikhsanov \cite{ihsan02} revived the propeller
model of Davies \& Pringle (1981) \cite{DP81} which incorporates 
$\dot{J}_{\ast} \propto -\omega_{\ast}$. This model
agrees with Eq.~(\ref{torque}) in the large $\omega_{\ast}$ limit. 
In the form we adopt, the torque entails
asymptotic approach to rotational equilibrium which is necessary for
addressing the period clustering of AXPs. The
recent numerical work \cite{romanova} also indicate an
efficient propeller torque ($\dot{J}_{\ast}\propto
-\Omega_{\ast}^{4/3}$). 

Being aware that modifications in $\alpha$ can be compensated by modifying
the initial inflow rate which in turn depends on the uncertain value of the
dynamical time scale, we simplify the EA03 model as follows:
\begin{equation}
\alpha =
\begin{cases}
5/4 & \text{if $\omega_{\ast}<1$;} \\
4/5 & \text{if $\omega_{\ast}>1$.}
\end{cases}
\label{alp}
\end{equation}
The value 5/4 is motivated by the accretor type of solution and 
the value 4/5 is motivated by the propeller type of solution. Both 
values correspond to the bound free opacity regime. The electron
scattering opacity prevails only in the $\sim 20$ years and 
does not effect the period evolution significantly.

\section{Results and Discussion}

We have followed the evolution of the star-disk system for
$10^{5}$ years, solving the torque equation (\ref{torque})
numerically with the adaptive step-size Runge-Kutta method
\cite{NR}.

\begin{figure}[ht]
\psfig{figure=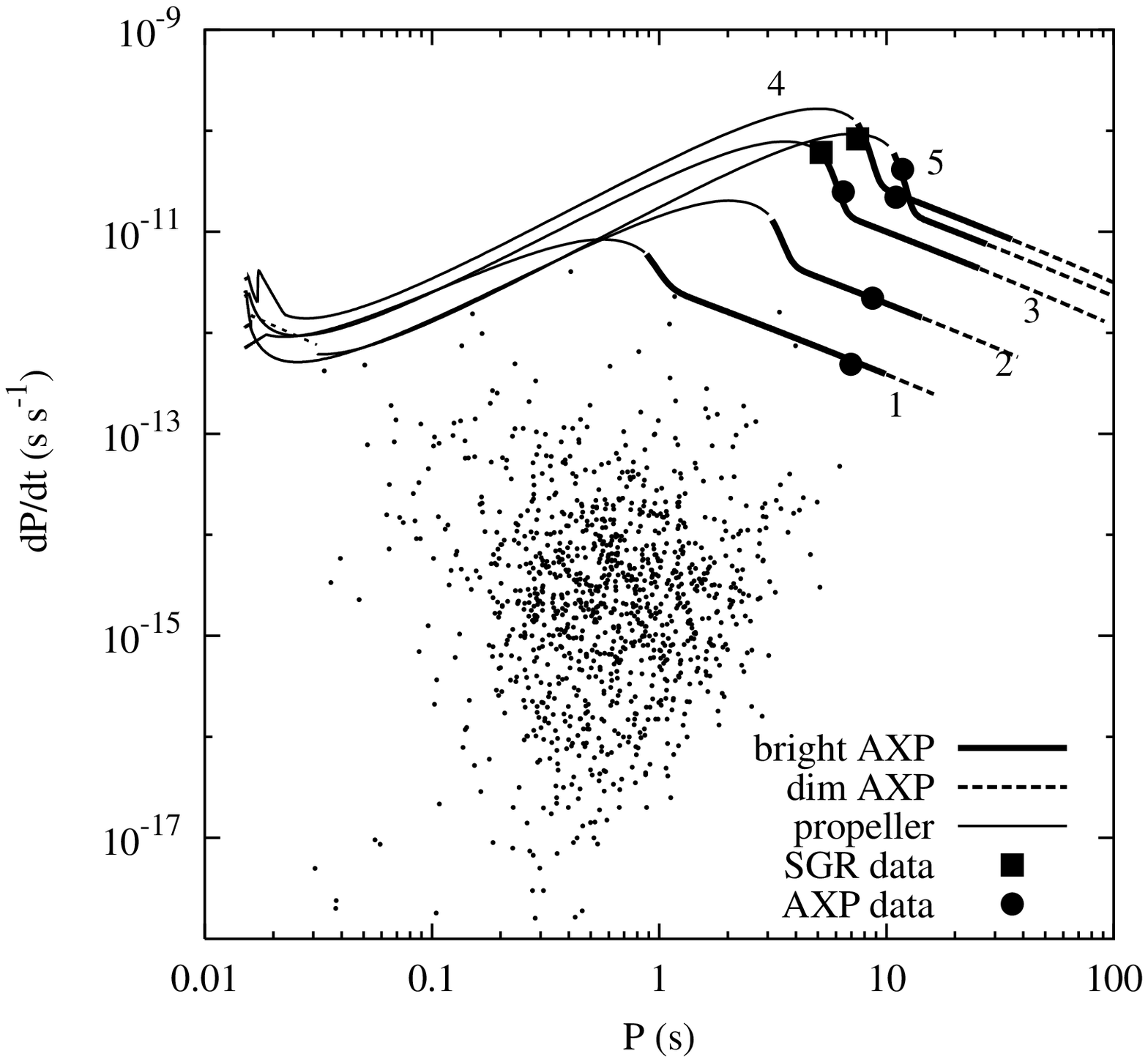, height=7cm, width=7cm}
\caption{Evolutionary tracks in the
$P-\dot{P}$ diagram covering all AXPs and two of the SGRs with known
period derivatives.
The \emph{thin solid  lines}
show the rapid spindown phase, the \emph{thick solid
lines} stand for the ``bright AXP'' stage and the \emph{dashed lines} show 
the ``dim AXP'' stage after the disk is neutralized ($\dot{M}<10^{16}$ g s$^{-1}$). 
The \emph{dots} denote the
radio pulsars. AXPs are shown by \emph{filled circles} and SGRs by
\emph{squares}. With different $\mu_{30}$ and $\dot{M}_0$, the model 
can cover all AXPs and SGRs.
The parameters for each track are as follows: 
(1)$B_{12}=1.8$, $\dot{M}_0=5\times 10^{28}$ g s$^{-1}$; 
(2)$B_{12}=2.8$, $\dot{M}_0=1\times 10^{28}$ g s$^{-1}$; 
(3)$B_{12}=5.5$, $\dot{M}_0=5\times 10^{27}$ g s$^{-1}$;
(4)$B_{12}=8$, $\dot{M}_0=3.2\times 10^{27}$ g s$^{-1}$;
(5)$B_{12}=6$, $\dot{M}_0=2\times 10^{27}$ g s$^{-1}$.
The initial period is 15 ms for all simulations.
\label{PPdot}}
\end{figure}

\begin{figure}[ht]
\psfig{figure=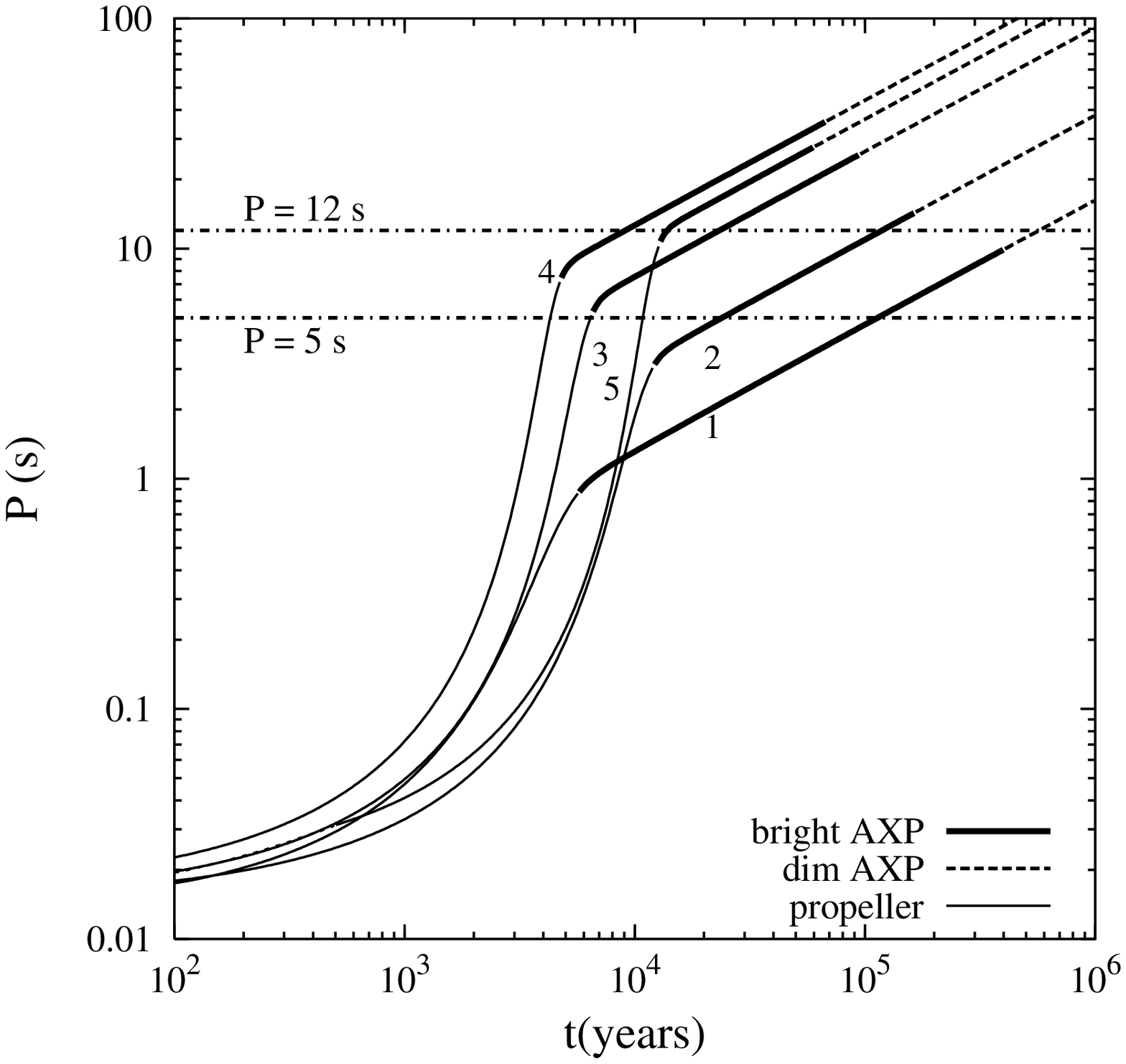, height=7cm, width=7cm} 
\caption{Period evolution of AXPs.
The \emph{thin solid lines} show the rapid
spindown phase, the \emph{thick solid lines} stand for the
``bright AXP'' stage and \emph{dashed lines} show the ``dim AXP'' stage 
after the disk is neutralized. The \emph{dashed and dotted horizontal lines} show the
period range of AXPs and SGRs. The parameters are the same as in Figure~(\ref{PPdot}) and
the numbers labelling the tracks correspond to the numbers of the tracks in Figure~(\ref{PPdot}).
\label{period}}
\end{figure}

In Figure~(\ref{PPdot}) we show five evolutionary tracks on the $P-\dot{P}$ diagram that
can cover seven AXP/SGRs for which period derivatives are available. 
Figure~(\ref{period}) shows the corresponding period evolutions for each of the tracks in
Figure~(\ref{PPdot}), tracks with the same numbers having the same parameters.
We assumed the radius and mass of the neutron star to
be $R_{\ast}=10^6$ cm and $M_{\ast}=1.4M_{\sun}$, respectively. 
Initial period is $P_0=15$ ms and we have assumed that the disk neutralizes 
when the mass flow rate drops below $10^{16}$ g s$^{-1}$ \cite{MPH01}.

Extending the 
analysis to disk evolution in the propeller regime
provides favorable effective power-law indices $\alpha_{\mbox{eff}}<1$.
However, in the availability of choosing a larger initial disk mass
one can always obtain AXPs even if the disk looses mass with $\alpha=5/4$. 

Considering the effect of the radiation pressure on the inner radius 
indicates a larger dynamical time scale, $T_0$. With a larger $T_0$, one can 
produce the same mass flow history with a smaller initial mass inflow rate corresponding
to a smaller initial disk 
mass. The numerical calculations (see \cite{pringle91}) indicate that
the power-law evolution corresponding to the self-similar solutions
start in a viscous time-scale which is still larger than the dynamical 
time scale. Given the ambiguity in $T_0$, estimations of initial disk
mass from power-law models discussed above, should not be taken very seriously.

Modifications in $\alpha$ and $T_0$ can always be compensated in the sense 
that one can obtain the same period evolution by tuning the initial 
mass flow rate. Such modifications then lead to the same mass flow rate
at the quasi-equilibrium stage at which the system is observed as an AXP.

The part of the model that is critical in determining whether fall-back disks
can produce AXPs or not is the propeller torque as first shown by \cite{FW}. 
If the torque model
is inefficient, one must then employ a high mass flow history to spin-down the 
neutron star to AXP periods which in turn results with very high luminosities 
in the quasi-equilibrium stage. As the torque is dependent on the inner radius
of the disk, the results are effected by the uncertain factors ($\xi=R_m/R_A$)
in determining the inner radius.

It is possible that a fall-back disk loses mass also in the propeller regime 
because of the propeller
effect itself, depending on the fastness parameter. The simple power-law models for the
flow rate are working models which do not
take into account the coupling of the star to the structure of the disk. 
A more
realistic treatment of fall-back disks would consider the 
effects of the magnetic field on
disk evolution, realistic opacities, including the iron line
opacity \cite{fryer} for the fall-back disk which is rich in
heavy elements, and the dependence of the accretion rate itself on
the motion of the inner radius 
$\dot{M}=2\pi R_m \Sigma(R_m)[\dot{R}_m-v_r(R_m)]$
(see \cite{spruit93}). 
All of these effects would result in time evolution of the disk that
cannot be described by simple power laws. Taking account of the
complexities of the real problem is not likely to produce
qualitative disagreement with the evolutionary timescales 
as long as the torque is efficient.

\end{document}